\documentclass[pre,twocolumn,amsmath,amssymb,superscriptaddress]{revtex4}
\usepackage{bm}
\usepackage[dvipdfm]{graphicx}
\usepackage{fancybox}
\usepackage{mathrsfs}
\usepackage[dvips]{color} 
\newcommand{\ltsimscript}{\protect\raisebox{-0.5ex}{$\stackrel{\scriptstyle >}{\sim}$}}
\newcommand{\gtsimscript}{\protect\raisebox{-0.5ex}{$\stackrel{\scriptstyle <}{\sim}$}}

\begin{document}

\title{Relevance of intracellular polarity to accuracy of eukaryotic chemotaxis}

\author{Tetsuya Hiraiwa}
\affiliation{Center for Developmental Biology, RIKEN, Chuo-ku, Kobe, Hyogo, Japan}
\author{Akihiro Nagamatsu}
\affiliation{Department of Mathematical and Life Sciences, Hiroshima University,1-3-1 Kagamiyama, Higashi-Hiroshima 739-8526, Japan}
\author{Naohiro Akuzawa}
\affiliation{Department of Mathematical and Life Sciences, Hiroshima University,1-3-1 Kagamiyama, Higashi-Hiroshima 739-8526, Japan}
\author{Masatoshi Nishikawa}
\affiliation{Center for Developmental Biology, RIKEN, Chuo-ku, Kobe, Hyogo, Japan}
\affiliation{Japan Science and Technology Agency, CREST, 1-3 Yamadaoka, Suita, Osaka 565-0871, Japan}
\author{Tatsuo Shibata}
\affiliation{Center for Developmental Biology, RIKEN, Chuo-ku, Kobe, Hyogo, Japan}
\affiliation{Department of Mathematical and Life Sciences, Hiroshima University,1-3-1 Kagamiyama, Higashi-Hiroshima 739-8526, Japan}
\affiliation{Japan Science and Technology Agency, CREST, 1-3 Yamadaoka, Suita, Osaka 565-0871, Japan}

\date{\today}

\begin{abstract}
Chemotactic cells establish cell polarity in the absence of external guidance cues. 
Such self-organized polarity is induced by spontaneous symmetry breaking in the intracellular activities,
which produces an emergent memory effect associated with slow-changing mode. 
Therefore, spontaneously established polarity should play a pivotal role in efficient chemotaxis. 
In this study, we develop a model of chemotactic cell migration that demonstrates the connection between intracellular polarity and chemotactic accuracy.
Spontaneous polarity formation and gradient sensing are described by
a stochastic differential equation. 
We demonstrate that the direction of polarity persists over a characteristic time 
that is predicted to depend on the chemoattractant concentration. 
Next, we theoretically derive the chemotactic accuracy as a function of both the gradient sensing ability and
the characteristic time of polarity direction.
The results indicate that the accuracy can be improved by the polarity.
Furthermore, the analysis of chemotactic accuracy 
suggests that accuracy is maximized at some optimal responsiveness to extracellular perturbations.
To obtain the model parameters, we studied the correlation time of random cell migration
in cell tracking analysis of \textit{Dictyostelium} cells. 
As predicted, the persistence time depended on the chemoattractant concentration.
From the fitted parameters, we inferred that polarized \textit{Dictyosteium} cells can respond optimally to a chemical gradient. Chemotactic accuracy was almost 10 times larger than can be achieved by non-polarized gradient sensing.
Using the obtained parameter values,
we show that polarity also improves the dynamic range of chemotaxis.
\end{abstract}

\maketitle

\section{Introduction}
Cell polarity maintains cell orientation 
by anisotropic distribution of particular factors. As such, it is an essential feature of many cells and tissues~\cite{Drubin96}.
For example, budding yeasts preserve the orientation of their budding site with respect to the original junction with their mother cell \cite{Drubin96}.
As another example, epithelial cells retain their apical and basal sides to establish and maintain sheet structures~\cite{Drubin96}.
Single migrating cells 
such as cellular slime mold \textit{Dictyostelium discoideum}, mammalian leukocytes and neurons
migrate in response to extracellular guidance cues. This directional cell migration is called chemotaxis.
Establishment and control of cell polarity is particularly important in eukaryotic chemotaxis, 
such as occurs in developmental processes, wound healing and immunological responses \cite{Friedl09}.
Thus, it is natural to question how cell polarity contributes to chemotaxis.

{\it Dictyostelium} is a well-studied model organism for investigating eukaryotic chemotaxis.
Its chemotaxis signaling pathways have been well characterized~\cite{Swaney10},
and its spontaneous and chemotactic migration behaviors 
have been extensively investigated~\cite{Fisher, Song06, Li08, Fuller10, Takagi, Li11,Boedeker10}. 
These studies have revealed  
high chemotactic accuracy of {\it Dictyostelium} cells 
even in shallow chemical gradients
and low chemoattractant concentrations~\cite{Fisher}.
The experimentally determined chemotactic accuracy is higher than expected,
assuming that the cell estimates the gradient direction as efficiently as possible.
Maximum likelihood estimates of the direction of orientation~\cite{Hu2010PRL} 
yield a chemotaxis index of $\mathrm{CI} \sim 0.023$ for these cells~\footnote{
The maximum likelihood estimation assumes \textit{Dictyostelium} cells
with $80,000$ receptors on the cell surface,  
dissociation constant $K_d= 180~\mathrm{nM}$,
and $1\%$ gradient with average chemoattractant concentration $C_0 = 0.25~\mathrm{nM}$.}.
However, Fisher's seminal experiment~\cite{Fisher}
yielded a chemotaxis index $\mathrm{CI}$ of $\mathrm{CI} \sim 0.12$ in a chemical gradient of $25~\mathrm{nM/mm}$.

\begin{figure}[t]
 \centering
 \includegraphics[width=8cm]{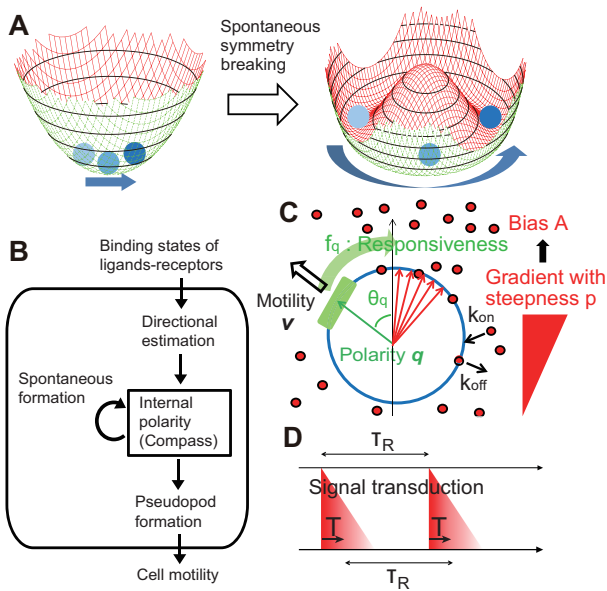}
 \caption{
Theoretical chemotaxis model with internal polarity and directional sensing.
(A) Spontaneous rotational symmetry breaking. 
(B) Schematic of signaling processing in eukaryotic chemotaxis. 
(C) Schematic of the presented model.
(D) Propagation of time delay in the signal transduction cascade.}
 \label{fig:introduction}
\end{figure}

\begin{figure*}[t]
 \centering
 \includegraphics[width=13cm]{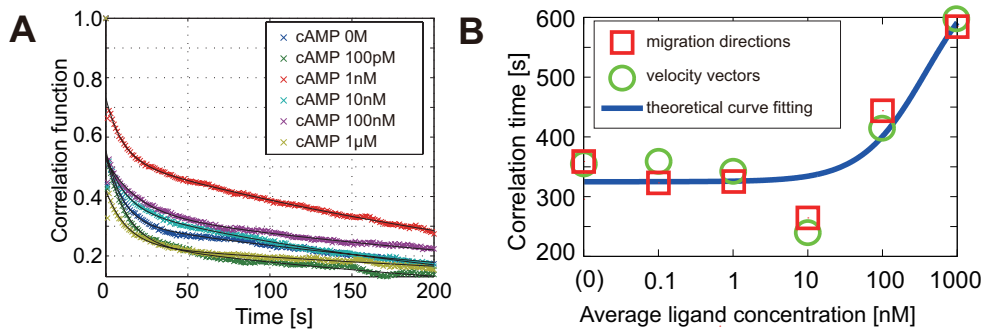}
 \caption{
Persistent cell migration in \textit{Dictyostelium} cells.
(A) Auto-correlation functions of the direction of cell migration. 
(B) Persistence time of cell migration as a function of average ligand concentration.
The correlation times of migrating directions (red squares) 
and velocity vectors (green circles) were experimentally obtained.
The solid line indicates Eq. \eqref{eq:tauvtauq},
fitted to the data indicated by the red squares.
The fitting parameters are obtained as 
$f_q \sim 0.0572 \ {\rm rad/s}$ and $ D^{\rm int.} \sim 0.00144 \  {\rm rad}^2/{\rm s}$.}
\label{fig:spontaneousTvsE}
\end{figure*}

Like many chemotactic eukaryotic cells, \textit{Dictyostelium} cells spontaneously polarize and migrate in random directions in the absence of extracellular cues.
This polarity
manifests as self-organization in signaling systems~\cite{Arai2010,Shibata2012,Taniguchi2013,Postma2004},
polarized structures in the cytoskeletal network, and localization of subcellular organelles.
To distinguish these polarities from specific cell-shapes and anisotropy in external environments,
we collectively label them as ``internal polarity''.
Spontaneous internal polarity in the absence of extracellular polarity implies 
a spontaneous breakdown of the internal isotropy or rotational symmetry.
Such spontaneous violation of a continuous symmetry can be illustrated by
a state point in the bottom of a potential shaped like a Mexican hat, as illustrated in the right panel of Fig.~\ref{fig:introduction}A.
This symmetry breaking is accompanied by a memory effect associated with 
a slowly-changing directional parameter.
In other words, in the asymmetric state, cells maintain their current direction.
By contrast, in the symmetric state,
directional information rapidly dissipates ~(see Fig.~\ref{fig:introduction}A left).
Asymmetric polarity 
is considered to underlie the persistence of cell migration
over a characteristic period of time, called the persistence time. 
The persistence time of \textit{Dictyostelium} cells ranges from $200$ to $600$ sec, depending on the chemoattractant concentration 
(Fig. \ref{fig:spontaneousTvsE}B),
much longer than the time constant of the chemoattractant-receptor reaction ($\sim 1$ sec).

Such spontaneous intracellular polarities are essential to high chemotactic accuracy.
Under shallow gradients, activities in the signal transduction network are modulated by
small changes in the likelihood of association between receptor and chemoligand,
which induce gradient responses. 
In the presence of homogeneously-distributed chemoattractant, 
the distribution of receptor occupancy along the cell periphery fluctuates due to its inherent stochasticity. 
Similar to external gradients, 
these spatial fluctuations can also drive the cell to orientate in a different direction~\cite{Tranquillo:1987tn}.
In fact, a short chemoattractant stimulus of 1 sec., which is similar to the time constant of receptor-ligand reaction~\cite{Ueda2001}, 
can modulate the state in the signaling pathway of gradient sensing in {\it Dictyostelium} cell \cite{Beta07}.
Therefore, we expect that random cell migration under uniform conditions  
is linked to the gradient response. 
Because the fluctuations in receptor occupancy depend on the chemoattractant concentration,
this concentration might also affect random cell migration.
By building these two quantities into a theoretical framework and testing the theory by experiment,
we can understand not merely how they are connected, 
but also elucidate how symmetry breaking contributes to chemotaxis in shallow chemical gradients.

Therefore, in this paper, we seek the connection between 
random cell migration in a uniform chemoattractant 
and chemotaxis in shallow gradients.
Several
theoretical models have been proposed to explain 
the persistence of cell migration~\cite{Takagi, Li11}.
Chemotactic cell migrations have also received theoretical treatment
\cite{Haastert07,Schienbein93,Schienbein94,Hu10}.
However, none of the existing theoretical models can explain chemotaxis in shallow gradients
or its relevance to random cell migration in a uniform chemoattractant. 
First, we study the persistence of random cell migration 
established by spontaneous internal polarity (Figs.~\ref{fig:introduction}B and C).
We found that the persistence time obtained from the correlation function of migration direction 
depends on chemoattractant concentration,
indicating a chemokinetic effect.
We then introduce directional bias by applying an external gradient (Figs.~\ref{fig:introduction}B and C). Considering that the internal polarity responsible for the persistence time is equally applicable to chemotaxis, 
we demonstrate that cell polarity enhances the chemotactic accuracy.
Finally, we determine the model parameters by comparing our theoretical results with experimental data of randomly migrating \textit{Dictyostelium} cells 
in a uniform condition.
We show that cell polarity can account for the
the high accuracy of chemotaxis
exhibited by \textit{Dictyostelium} cells
in Fisher's study {\it et al.} \cite{Fisher}.

\section{Results} \label{sec:spontaneousTheory}
\subsection{Random cell migration under uniform conditions}

We first consider cell migration under uniform conditions, {\it i.e.} in the absence of a chemoattractant gradient. 
As mentioned above, we consider a spontaneously maintained internal polarity in the intracellular process. 
The cells migrate in the direction of the internal polarity, which is subject to change by intra- and extra-cellular perturbations.
For simplicity, we assume that the magnitude of the internal polarity and the migration speed are constant.

Intracellular perturbations originate from stochastic variability of signal reactions and other processes.
Such perturbations introduce randomness to the internal polarity, 
leading to random cell migration. 
The random orientation of the polarity direction~$\theta_q$ can be described by 
\begin{equation}
\frac{d}{dt}\theta_q(t)=\xi(t)
\end{equation}
where $\xi(t)$ is a random perturbation 
described by white Gaussian noise with $\langle\xi(t)\rangle=0$
and $\langle\xi(t)\xi(t')\rangle=2D\delta(t-t')$.
In the absence of extra-cellular perturbation, 
the total noise strength is $D=D^{\rm int.}$ (rad$^2$/sec), where $D^{\rm int.}$ is the dispersion strength of intracellular stochastic perturbations.

When the cell is exposed to a spatially uniform chemoattractant, the stochastic binding of receptors and chemoattractant ligands constitutes another noise source. 
The information of the binding states at a particular time 
decays exponentially with time constant $\tau_R$.
Here, $\tau_R$ is the correlation time of the receptor state, given by \cite{Hu10}
\begin{equation} \label{eq:tauR}
\tau_R(C_0)=\frac{1}{k_d + k_a C_0}=\frac{1}{k_d ( 1+ K_d^{-1} C_0 )} \ ,
\end{equation}
where $k_d$ and $k_a$ are the dissociation and association rates, respectively, and $K_d=k_d/k_a$ is the dissociation constant.
We denote the average impact of individual changes in the binding state on the internal polarity by $f_q$.
The integrated noise strength during time interval $\tau_R$ scales as $f_q \tau_R$.
Since the noise dispersion is proportional to the square of the noise strength, 
the dispersion of the extracellular-noise contribution $D^{\rm ext.}$ through receptors 
is estimated as $D^{\rm ext.} \sim (f_q \tau_R)^2 \times N^2 \times (1/N)$, 
where $N \sim 1/\tau_R$ is the number of independent events per unit time.
Thus, the dispersion is given by 
\begin{equation}
D^{\rm ext.}=\frac{f_q^2 \tau_R}{2} \ .
\label{eq:D_ext}
\end{equation}
The numerical factor, $1/2$, is given by the detailed calculation in Materials and Models (Eq. \eqref{eq:xiMSD}).

While internal polarity dictates the moving direction of the cell, the direction of the internal polarity fluctuates stochastically under intra- and extra- cellular perturbations, as explained above.
Therefore, with the resulting increase in total dispersion, $D=D^{\rm ext.}+D^{\rm int.}$,
the correlation time $\tau_c$ of the migration direction or the internal polarity should decrease.
Given that the angular diffusion constant $D$ has an inverse time unit, 
the correlation time $\tau_c$ is given by $\tau_c = D^{-1}$.
For a detailed derivation of this relation, see Eq. \eqref{eq:autocorrelation} in Materials and Models.
Thus, from Eqs. \eqref{eq:tauR} and \eqref{eq:D_ext}, 
the correlation time $\tau_c$ is given by
\begin{align} 
 \tau_c(C_0) &= (D^{\rm ext.}+D^{\rm int.})^{-1} \notag \\
&= \left( \frac{f_q^2}{2 k_d(1 + K_d^{-1} C_0)} + D^{\rm int.} \right)^{-1} \ . \label{eq:tauvtauq}
\end{align}
This equality (Eq. \eqref{eq:tauvtauq}) indicates that
the motile behavior of cells depends upon chemoattractant concentration,
even if the chemoattrantant is homogeneously distributed in space.
The correlation time $\tau_c$ increases with increasing average concentration $C_0$,
as shown in Fig. \ref{fig:spontaneousTvsE}B (blue solid line).
By fitting Eq. \eqref{eq:tauvtauq} to the measured concentration dependence of the correlation time, and knowing the values of the receptor parameters,
we can estimate the values of parameters $f_q$ and $D^{\rm int.}$.

\subsection{The effect of internal polarity on the gradient response and chemotactic accuracy}
In the presence of chemoattractant gradients,
cells perform gradient sensing, and
orient their internal polarity in the direction of the gradient. 
The quality of the gradient sensing may depend on the stochasticity in the receptor states. 
When the noise increases or decreases relative to the signal strength, 
the accuracy of gradient sensing may decrease or increase, respectively~\cite{Ueda08, Amselem2012}.
Thus, the effective gradient information,
which we quantify by a non-dimensional parameter $A$,
may decrease if the stochastic noise increases.
This bias $A$ also depends on the chemoattractant concentration and the steepness of the gradient.

The response of cells to gradient perturbations 
should depend on the responsiveness $f_q$ (rad/sec), introduced above. 
As $f_q$ increases, 
the chemoattractant gradient should exert greater effect.
Thus, the "driving force" $S$ exerted on the internal polarity depends on both the bias and the responsiveness, which may be expressed as 
\begin{equation}
S=\frac{f_q A}{2}.
\label{eq:force}
\end{equation}
(For details of the derivation, see Eq. \eqref{eq:paverage}.)
\begin{figure*}[!t]
 \centering
 \includegraphics[width=13cm]{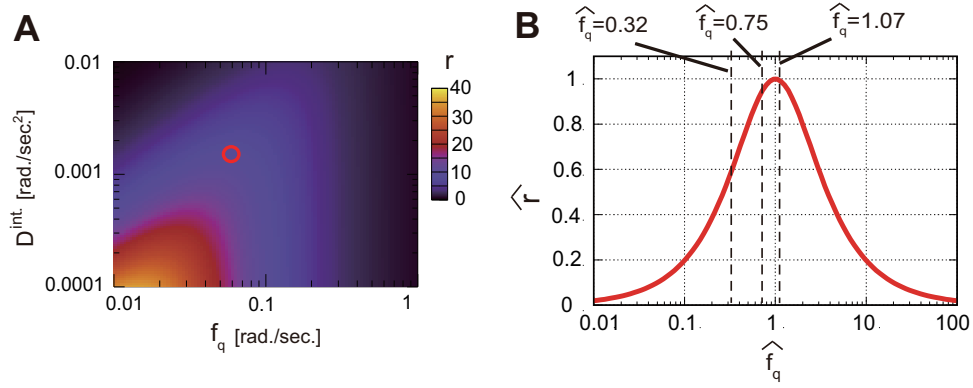}
 \caption{
The effect of internal polarity on chemotactic accuracy.
(A) The amplification ratio $r$ is given by Eq.~\eqref{eq:amplificationratio-orig} 
 in the low concentration limit $C_0 \rightarrow 0$ for $k_d=1 \ {\rm s}^{-1}$.
 The red circle indicates the parameters of {\it Dictyostelium} cells, namely,
 $f_q \sim 0.0572 \ {\rm rad/s}$ and $ D^{\rm int.} \sim 0.00144 \  {\rm rad}^2/{\rm s}$.
(B) The scaled amplification ratio $\hat{r}$ given by Eq.~\eqref{eq:amplificationratio}.
 Horizontal axis is the dimensionless responsiveness $\hat{f}_q= f_q /\sqrt{2D^{\rm int.}/\tau_R}$.
 The dimensionless responsiveness for 
 the given parameters, estimated from the correlation time of spontaneous migration
 (see Eqs.~\eqref{eq:fittingfq} and \eqref{eq:fittingDI}),
 is about $1.07$ in the low average concentration limit $C_0 \rightarrow 0$,
 $0.753$ at $C_0=180 \ {\rm nM}$, and $0.321$ at $C_0=1800 \ {\rm nM}$.
}
 \label{fig:amplificationfactor}
\end{figure*}

Fisher {\it et al.} \cite{Fisher}
specifies the accuracy $\kappa$ by 
the sharpness of the distribution $P(\theta_v)$ of migration directions $\theta_v$, defined by $P(\theta_v) \propto \exp(\kappa \cos \theta_v)$.
The chemotaxis index, defined as the statistical average of the migration direction cosines, is ${\rm CI} \equiv \langle \cos \theta_v \rangle$. 
When the chemotactic accuracy $\kappa$ is small, the CI
is roughly $\kappa/2$ \cite{Fisher}.
$\kappa$ is also expected to be proportional to the bias $S$.
Furthermore, $\kappa$ should improve 
as the stochastic perturbations $D^{\rm ext.}+D^{\rm int.}$ decrease,
or conversely, as the correlation time $\tau_c$ increases. 
These conditions are satisfied if $\kappa$ takes the form
\begin{equation} \label{eq:kappa}
 \kappa = \tau_c S = \frac{A}{f_q k_d^{-1} (1 + K_d^{-1} C_0)^{-1} + 2 f_q^{-1} D^{\rm int.}} \ .
\end{equation}
(Details of the derivation are provided in Materials and Models; see Eq. \eqref{eq:stationarydistribution}).

To see the significance of the internal polarity on the chemotactic accuracy,
we consider the ratio between the chemotactic accuracy $\kappa$ and the bias $A$, denoted as the amplification ratio $r \equiv {\kappa}/{A}$.
This ratio quantifies the extent to which the internal polarity enhances the chemotactic accuracy.
From Eq. \eqref{eq:kappa}, the amplification ratio $r$ is obtained as 
\begin{equation} \label{eq:amplificationratio-orig}
 r = \frac{1}{f_q \tau_R + 2 f_q^{-1} D^{\rm int.}} \ ,
\end{equation}
which can be rescaled as 
\begin{equation} \label{eq:amplificationratio}
 \hat{r} \equiv \frac{r}{\sqrt{1/(8\tau_R D^{\rm int.})}} = \frac{2}{\hat{f_q} + \hat{f_q}^{-1}},
\end{equation}
where $\hat{f_q} = f_q/\sqrt{2D^{\rm int.}/\tau_R}$ is the dimensionless responsiveness.
In Fig.~\ref{fig:amplificationfactor}A 
the amplification ratio $r$ is plotted as a function of $f_q$ and $D^{\rm int.}$ for $\tau_R=1$ second.
Figure~\ref{fig:amplificationfactor}B plots 
the rescaled amplification ratio $\hat{r}$ 
as a function of the rescaled responsiveness $\hat{f_q}$. From this figure, we 
observe that the amplification is maximized ($\hat{r}=1$) at $\hat{f_q}=1$.
These results suggest that, for a given value of $D^{\rm int.}$, there exists 
an optimum responsiveness $f_q$ that maximizes the chemotactic accuracy.
This maximum accuracy is achieved when 
the external and intrinsic perturbations are balanced.
If the responsiveness $f_q$ is small,
the correlation time~$\tau_c$ in Eq. \eqref{eq:tauvtauq} is almost independent of~$f_q$,
because the dispersion is dominated by the intrinsic stochastic perturbation $D^{\rm int.}$.  
Simultaneously, the external driving force $S$ in Eq. \eqref{eq:force} is strengthened as $f_q$ increases.
Therefore, 
for small responsiveness $f_q$,
the accuracy $\kappa$ increases proportionally to the responsiveness $f_q$.
Conversely, if the responsiveness $f_q$ is large,
the correlation time is dominated by extrinsic stochastic perturbations ~$\tau_c$ as $\tau_c\propto f_q^{-2}$ in Eq. \eqref{eq:tauvtauq}.
In this case, the accuracy $\kappa$ in Eq. \eqref{eq:kappa} decreases in proportion to $f_q^{-1}$.

\subsection{Chemokinesis during random cell migration of \textit{Dictyostelium} cells} 
Here, we determine the values of the parameters~$f_q$ and $D^{\rm int.}$ of \textit{Dictyostelium} cells. To this end, we
study the dependence of 
the correlation time of the migration direction on the extracellular concentration of the chemoattractant cAMP. As we have mentioned, 
the stochastic binding of chemoattractant by the receptor 
contributes additional noise to the internal polarity. 
Because the characteristic time of the receptor depends on the chemoattractant concentration as shown in Eq.~\eqref{eq:tauR}, 
the correlation time of the migration direction should depend on the chemoattractant concentration, as predicted in 
Eq.~\eqref{eq:tauvtauq}.
To investigate this expected phenomenon, we tracked the migration of single \textit{Dictyostelium} cells in uniform cAMP concentrations (see Materials and Models).

We first obtained the trajectories of cell centroids at 1 s intervals for longer than 5 min up to 30 min.
The velocity vector of the cell centroid, $\boldsymbol{v}(t)=v(\cos\phi(t),\sin\phi(t))$, was determined from
the temporal autocorrelation function~$C(t)$ of the migration direction $\phi(t)$, 
given by $C(t)=\langle\cos(\phi(t)-\phi(0))\rangle$. 
Figure \ref{fig:spontaneousTvsE}A plots $C(t)$ at different cAMP concentrations.
The correlation function~$C(t)$ can be fitted by a sum of two exponential functions; 
\begin{equation}
C(t)=C_1e^{-t/\tau_1}+C_2e^{-t/\tau_2}.
\end{equation}
The shorter time constant $\tau_1$, characterizing rapid deformations in the cell shape, is 10-20 sec, while the 
longer one $\tau_2$, which quantifies the persistence time in the migration direction, is several hundred seconds.
Here we focus on the time scale of the persistence; therefore,
we compare $\tau_2$ with the correlation time $\tau_c$ in Eq. \eqref{eq:tauvtauq}.
As shown in Fig.~\ref{fig:spontaneousTvsE}B,
the time constant~$\tau_2$ increases as the cAMP concentration increases. 
We note that the correlation times of the centroid velocity~
$\boldsymbol{v}(t)$ and the direction~$\phi(t)$ essentially correspond with each other (Fig. \ref{fig:spontaneousTvsE}B).
This indicates that the constant speed approximation used in Eq. \eqref{eq:tensorv} is valid in this time scale.
\begin{figure}[!t]
 \centering
 \includegraphics[width=8cm]{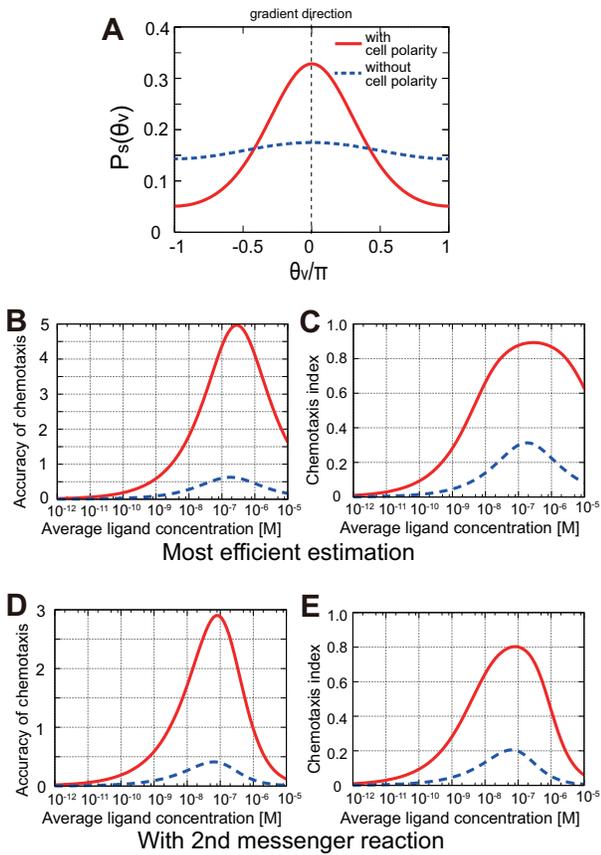}
 \caption{
Chemotactic accuracy.
(A) Stationary distribution of the migration direction $P_s(\theta_v)$
with $\theta_v=\phi$ in the absence of cell polarity (blue broken line) 
and $\theta_v=\theta_q$ with cell polarity (red solid line). Parameters are:
 $A=0.1$, $f_q \sim 0.0572 \ {\rm rad/s}$, $ D^{\rm int.} \sim 0.00144 \  {\rm rad}^2/{\rm s}$ (obtained for \textit{Dictyostelium} cells as Eqs. \eqref{eq:fittingfq} and \eqref{eq:fittingDI}.). 
 (B-C)  (B) The accuracy of chemotaxis $\kappa$ and (C) the chemotaxis index $\textrm{CI}$
plotted as functions of 
average chemoattractant concentration~$C_0$.
At all concentrations, the gradient steepness is $p=0.01$.
The chemotactic accuracy~$\kappa$ in \eqref{eq:kappa} 
and chemotaxis index $\textrm{CI}$ in \eqref{eq:CIresult} 
achieved with internal polarity
are indicated by the red solid lines,
while those without internal polarity (calculated from
Eq. \eqref{eq:CIWithoutPolarity}; $\kappa=A$) 
are indicated by the blue broken lines.
The bias $A$ was determined from Eq. \eqref{eq:A0extapprox} together with Eq. \eqref{eq:muwithout}.
(D-E) 
The effect of intracellular noise in the second messenger reaction
in (D) the accuracy of chemotaxis $\kappa$ and (E) the chemotaxis index $\textrm{CI}$, as functions of ligand concentration.
The bias $A$ was determined from Eq. \eqref{eq:A0extapprox} together with Eq. \eqref{eq:muwith}.
}
 \label{fig:chemotaxisindex}
\end{figure}

To obtain the values of $f_q$ and $D^\mathrm{int.}$,
we fitted Eq.~\eqref{eq:tauvtauq} to the time constants 
obtained for the centroid direction~$\phi$
with the known receptor constants.
As shown by the solid curve in Fig.~\ref{fig:spontaneousTvsE}B,
the dependence of $\tau_2$ on the cAMP concentration can be well fitted by Eq.~\eqref{eq:tauvtauq}, with fitting parameters 
\begin{equation} \label{eq:fittingfq}
 f_q \sim 0.0572 \ {\rm rad/s}
\end{equation}
and 
\begin{equation} 
\label{eq:fittingDI}
 D^{\rm int.} \sim 0.00144 \ {\rm rad}^2/{\rm s} \ . 
\end{equation}

\subsection{Optimality of \textit{Dictyostelium} chemotaxis}

Using the parameter values specified by Eqs. \eqref{eq:fittingfq} and \eqref{eq:fittingDI},
we first studied the stationary distribution of the migration direction.
In Fig. \ref{fig:chemotaxisindex}A,
the distributions of migration direction
are plotted for $A=0.1$
in the absence and presence of internal polarity (Eq.~\eqref{eq:Pphi1} (blue broken line) and Eq. ~\eqref{eq:stationarydistribution} (red solid line), respectively). Clearly, the distribution of migration direction 
steepens around the gradient direction in the presence of internal polarity.

We next studied the dependence of chemotactic accuracy $\kappa$ and 
chemotaxis index ${\rm CI}$ on the chemoattractant concentration~$C_0$.
As shown in Fig.~\ref{fig:chemotaxisindex}B, 
in the presence of internal polarity, both the accuracy and dynamic range of chemotaxis, 
given by Eq.~\eqref{eq:kappa} (red solid line),
are larger than when internal polarity is absent
(blue broken line).
Together, these results indicate 
that internal polarity contributes not only to chemotactic accuracy but also to the
dynamic range of chemotaxis. 
Figure~\ref{fig:chemotaxisindex}C shows the dependence of the chemotaxis index ${\rm CI}$ 
on the concentration $C_0$.
The stochastic signal processing and transduction during chemotactic signaling 
has been investigated theoretically \cite{Ueda08} and experimentally \cite{Amselem2012}.
These previous studies showed that 
the relationship between chemotactic accuracy and extracellular cAMP concentration mirrors the signal-to-noise ratio at the level of the second messenger  ~\cite{Ueda08,Amselem2012}.
Therefore, here, we introduce intracellular noise in the second messenger reaction
to the bias~$A$ (see  
Eqs. \eqref{eq:A0extapprox}-\eqref{eq:muwithout} in Materials and Models), and investigate its effect.
The resultant chemotactic accuracy $\kappa$ and chemotaxis index ${\rm CI}$ as functions of cAMP concentration are 
shown in Figs.~\ref{fig:chemotaxisindex}D and E, respectively.
Again, both the accuracy and dynamic range of chemotaxis are improved by internal polarity.
Comparing Fig.~\ref{fig:chemotaxisindex}D with B,
we find that the accuracy $\kappa$ in the high concentration region $C_0 \ltsimscript K_d$ is diminished by stochasticity in the second messenger reaction.
In Fig.~\ref{fig:chemotaxisindex}D, both the ligand concentration of highest chemotactic accuracy and the dynamic range of chemotaxis agree 
with the experimental result of Fisher {\it et al.} \cite{Fisher}.

The effect of internal polarity on chemotaxis can be characterized by 
the amplification ratio $r$ introduced in Eq. \eqref{eq:amplificationratio-orig},
which defines the ratio 
of the chemotactic accuracies~$\kappa$ 
with and without internal polarity.
The amplification ratio $r$
depends on the values of $f_q$ and $D^\textrm{int.}$.
In Fig.~\ref{fig:amplificationfactor}A,
$r$ given by Eq.~\eqref{eq:amplificationratio-orig} is plotted 
in the low concentration limit $C_0 \rightarrow 0$, {\it i.e.} $\tau_R = 1$ second.
For the values of $f_q$ and $D^\mathrm{int.}$ of 
\textit{Dictyostelium} cells, indicated by the red circle in Fig.~\ref{fig:amplificationfactor}A,
the amplification factor is $r \sim 9.32$.
The amplification ratio $r$ depends on both $D^\mathrm{int.}$ and $C_0$ through $\tau_R$. 
To remove these dependences, 
we study the scaled amplification ratio $\hat{r}$ 
given by Eq.~\eqref{eq:amplificationratio}, which depends only on the scaled responsiveness  
$\hat{f_q} = f_q/\sqrt{2D^{\rm int.}/\tau_R}$
(plotted in Fig.~\ref{fig:amplificationfactor}B).
The scaled responsiveness of \textit{Dictyostelium} cells
is estimated as  
$\hat{f_q} \sim 1.07$, $0.753$, and $0.321$
for $C_0 \rightarrow 0$, $C_0 = K_d= 180~\text{nM}$, and $C_0 = 10\times K_d= 1800~\text{nM}$, 
respectively (see Fig.~\ref{fig:amplificationfactor}B, broken lines).

For $C_0 \gtsimscript K_d$,
the responsiveness is distributed around 
the maximum amplification ratio, 
indicating that the responsiveness parameter $f_q$ of \textit{Dictyostelium} cells is almost optimal in this concentration range.

\section{Discussion} 
\subsection{Chemotaxis ability of mutant cells}
The reduced chemotactic ability of mutant cells has been extensively studied.
From our proposed theory, 
we can identify potential chemotactic influences that are impaired by mutations.
In Eq. \eqref{eq:kappa}, 
the mutable parameters are the bias $A$,
the responsiveness~$f_q$ and the internal noise $D^{\mathrm{int.}}$.
A mutation can cause a reduction in the bias $A$,
an increase in the internal noise $D^\mathrm{int.}$, and/or a decline in the responsiveness~$f_q$.
Since an optimal $f_q$ exists in wild type cells (see Fig. \ref{fig:amplificationfactor}), 
decreasing the responsiveness $f_q$
will cause a decrease in chemotactic ability \footnote{Furthermore, 
if the chemoattractant concentration $C_0$ is smaller than $K_d$, either increasing or decreasing the responsiveness $f_q$ will cause the chemotactic ability to decrease.}.
Among these three parameters,
$f_q$ and $D^\mathrm{int.}$ affect the correlation time $\tau_c$ of 
random cell migration in isotropic conditions. 
By studying the dependence of this time constant $\tau_c$ on the chemoattractant concentration, 
we can identify which of $f_q$ or $D^\mathrm{int.}$ is modulated.
We also notice that amplification ratio $r$ is insensitive to
change of responsiveness $f_q$ around its optimal value
(see Eq. \eqref{eq:amplificationratio}).
Hence, the chemotactic accuracy of {\it Dictyostelium} cells
should not be sensitive to small changes in the responsiveness $f_q$, as shown in
Fig.~\ref{fig:amplificationfactor}B.

The directional correlation of random cell migration is characterized by ``persistence'', 
defined as the ratio of net displacement to total path length in a given time interval, is here denoted by $\mathcal{P}$ (it is also called ``directionality''). Note that $\mathcal{P}$ depends on the observational time interval $t$, and is independent of cell migration speed.
$\mathcal{P}$ and $\tau_C$ are related as follows;
In the absence of a chemoattractant gradient,
the persistence~$\mathcal{P}$ is almost $1$ when $t \ll \tau_c$. When $t \gg \tau_c$, it becomes
proportional to the square root of the correlation time; that is, 
$\mathcal{P}\sim\sqrt{\tau_c/t}$.
In the presence of a chemoattractant gradient,
$\mathcal{P} \sim 1$ when $t \ll \tau_c$,
$\mathcal{P} \sim \sqrt{\tau_c/t}$ when $\tau_c \ll t \ll \tau_A$,
and $\mathcal{P} \sim {\rm CI}$ when $t \gg \tau_A$.
Here, $\tau_A$ is the time scale over which 
the displacement induced by the gradient bias $\textrm{CI}\times v_0t$
dominates the diffusion length by randomness in the cell migration, given by~$v_0 \sqrt{\tau_C t}$.
Thus, $\tau_A$ is given by $\tau_A \sim {\rm CI}^{-2} \tau_C$.

The chemotactic ability of PI3K mutant of {\it Dictyostelium} cells has been studied in \cite{Takeda07,Bosgraaf08}.
PI3K is involved in the self-organization of the phosphatidylinositol lipids signaling system~\cite{Arai2010,Shibata2012},
and is a candidate constituent of the internal cell polarity.
In addition, PI3K is activated by RasG~\cite{Swaney10}.
RasG is involved in one of the parallel chemotactic signaling pathways in {\it Dictyostelium} cells~\cite{Swaney10}.
Therefore, we expect that 
$D^\mathrm{int.}$ is increased and $f_q$ is decreased in the PI3K mutant.
This may lead to lowering both the chemotactic accuracy in shallow gradients, as shown by Eq. \eqref{eq:amplificationratio},
and the persistence $\mathcal{P}$, regardless of whether chemoattractant is present.
These speculations have been verified by experimental observation \cite{Takeda07,Bosgraaf08}.

If the ability to detect and respond to the chemoattractant gradient are intact ({\it i.e.} the bias $A$ and responsiveness $f_q$ are intact), 
Eq. \eqref{eq:kappa} predicts that the ratio of chemotaxis index to the square of the persistence, {\it i.e.}, ${\rm CI}/\mathcal{P}^2$,
is approximately constant, with $\mathcal{P}$ defined in the interval $\tau_c \ll t \ll \tau_A$. 
This tendency has indeed been verified in PTEN mutant cells, in which 
CI/$\mathcal{P}^2$ was similar to that of wild type cells
(using data reported in Ref.~\cite{Wessels07}).
PTEN is also involved in the self-organization of the phosphatidylinositol lipid signaling system \cite{Arai2010,Shibata2012}.
This suggests that the loss of PTEN activity amplifies the intrinsic fluctuations $D^\mathrm{int.}$ 
of the polarity, while little affecting $A$ and $f_q$.
In support of this idea, lateral pseudopod formation is enhanced in PTEN mutant cells \cite{Wessels07}.
However, our assumption that the migration and polarity directions are linked may not be valid for PTEN mutants.
If this assumption is removed,  
the chemotaxis accuracy and persistence decrease.

The synergistic effect of several chemotaxis pathway has been studied in Ref.~\cite{Veltman08}. 
Simultaneous inhibition of both PI3K and sGC pathways induces a dramatic reduction 
in chemotactic ability ${\rm CI}$ and persistence $\mathcal{P}$, 
while inhibitions of either pathway cause a mild reduction, indicating the pathways exert a synergistic effect.
Although variations arise in both chemotactic ability and persistence, 
their ratios CI/$\mathcal{P}^2$ in the mutation and inhibition experiments of Ref.~\cite{Veltman08} appear to be
almost constant. This might indicate that neither gradient detection ability $A$ nor responsiveness $f_q$ are 
strongly affected by the mutation, and that the sGC pathway is involved in stabilizing the internal polarity $D^{\rm int.}$
via synergetic effects with the PI3K pathway.

\subsection{Other contributions relevant to chemotactic accuracy }

Several factors that may reduce chemotactic accuracy have been excluded from the model. 
The experimental accuracy of chemotaxis obtained 
by Fisher {\it et al.} \cite{Fisher}
is smaller than that obtained 
in our theory (see Fig.~\ref{fig:chemotaxisindex}D),
indicating that some of these factors are significant.

To obtain the maximum feasible accuracy, we assumed that all $80,000$ receptors are located at the periphery of the cell.
However, in reality, the receptors in the vicinity of the vertical point cannot contribute to determining the migration direction.
This effect reduces the effective number of receptors and hence decreases the chemotactic accuracy.

Spatial and temporal stochastic variations in the distribution of receptors and intracellular signaling molecules 
may also affect the dispersion in the chemotactic accuracy.
For example, as stated in Materials and Models, we have neglected the correlations
in the spatial density fluctuations of receptors and the activated second messengers
at different times and positions along the cell membrane.
However, these spatiotemporal stochastic fluctuations and correlations are expected because the receptors and activated second messengers can diffuse. 
In addition, the activation time is also subject to stochastic delays.
An intriguing future problem is to theoretically investigate the consequences of spatial and temporal correlations on the density fluctuations of activated second messengers.
Such studies might elucidate how cells overcome these noises.

We also postulated that the migration direction $\theta_v$ adiabatically follows the 
polarity direction $\theta_q$ on the time scale of interest.
However, 
$\theta_v$ may deviate from $\theta_q$ 
over time scales exceeding the correlation time of polarity direction. 
These deviations, which would further contribute to fluctuation in the motile direction, are observed in cell deformation processes.
Previously, we have reported that cell deformation can alter the gradient sensing ability, with subsequent effect on  
the probability distribution of chemotactic migration directions \cite{BabaPre, Hiraiwa13}.
Therefore, the temporal fluctuations observed in {\it Dictyostelium} cells 
may also affect the dispersion of motile directions.

\subsection{Outlook}
The present model, which incorporates internal polarity and gradient sensing, is applicable to several different cases, 
including chemotaxis with cell shape deformation.
In particular, slight deviations of the motile direction distribution from the normal circular distribution
can be explained by the present model,
provided that the process of cell deformation and the influence of cell shape on the gradient sensing are included \cite{Hiraiwa13}.
The current model can be used to study chemotaxis towards time-varying gradients, such as chemoattractant waves.
By introducing cell-cell interactions into the model, the collective chemotaxis of cell populations could also be investigated.
Furthermore, this kind of modeling should be applicable to the motility analysis of other kinds of chemotactic eukaryotic cells.

\section*{Materials and Models}
\subsection*{Mathematical modeling of internal polarity}

We consider the spontaneous formation of the internal polarity~$\boldsymbol{q}=(q_x,q_y)$, 
incorporating gradient sensing and internal and external perturbations, as described in Results.
The simplest evolution equation that describes 
such internal polarity~$\boldsymbol{q}=(q_x,q_y)$ 
is given by~\cite{Hiraiwa13}
\begin{equation} \label{eq:tensorq}
 \frac{d}{dt} \boldsymbol{q} = I_q ( 1 - |\boldsymbol{q}|^2) \boldsymbol{q} 
+ \boldsymbol{\xi}^\textrm{int.}(t) + \boldsymbol{\xi}^{\rm ext.}(t) \ .
\end{equation}
The first term on the right hand side describes  
the spontaneous symmetry breaking of isotropy
with polarity strength~$I_q$.
For simplicity, we assume $I_q \rightarrow \infty$, so that 
the internal polarity $\boldsymbol{q}$ becomes
$\boldsymbol{q}(t)=(\cos\theta_q(t),\sin\theta_q(t))$
with $|\boldsymbol{q}|=1$.
The second term $\boldsymbol{\xi}^{\textrm{int.}}(t)$ 
describes the internal noise,
assumed as Gaussian white noise
with $\langle {\bm \xi}^{\rm int.} \rangle =0$, and 
\begin{equation}
\langle \xi_i^{\rm int.}(t) \xi_j^{\rm int.}(t') \rangle = 2 D^{\rm int.} \delta_{ij} \delta(t-t').
\label{eq:xiintdispersion}
\end{equation}
The last term~$\boldsymbol{\xi}^{\textrm{ext.}}(t) \equiv f_q \boldsymbol{e}(t)$  
describes the gradient sensing
with responsiveness~$f_q$.
The unit vector $\boldsymbol{e} (t)$ 
specifies the inferred direction of the external gradient,~$\phi(t)$,
{\it i.e.} $\boldsymbol{e}(t)=(\sin \phi(t), \cos \phi(t))$.
Here, we consider that 
the cell estimates the extracellular gradient direction 
from the distribution of its chemoattractant-occupied receptors (see below).
In this paper, 
we consider cell motions 
occurring over the persistence time of cell migration.  
On these time scales, 
the velocity~$\boldsymbol{v}$ of the cell immediately follows the internal polarity as 
\begin{equation} \label{eq:tensorv}
 {\bm v} = v_0 {\bm q} \ ,
\end{equation}
where the migration speed $v_0$ is assumed constant as previously reported~\cite{Hu10}.

For a given spatial distribution of chemoattractant-occupied receptors, we consider the most probable estimate of the gradient direction, $\phi$.
The simplest probability distribution $P(\phi)$ of the estimated direction, assuming circular shaped cells,
is given by \cite{BabaPre}
\begin{equation} \label{eq:Pphi1}
 P(\phi) = \frac{1}{2\pi} + \frac{A}{2\pi} \cos \phi \ ,
\end{equation}
where the true gradient direction is~$\phi=0$.
Here, $A$ is the bias strength imposed by the external gradient. $A$ is presented in detail below.
In the absence of internal polarity, 
the chemotaxis index gained from directional inference alone is given as 
\begin{equation}\label{eq:CIWithoutPolarity}
\textrm{CI}=\int_0^{2\pi} d\phi \cos \phi P(\phi) = A/2.
\end{equation}

The characteristic time $\tau_R$ in Eq. \eqref{eq:tauR}
is much smaller than
the persistence time 
of $300$ seconds \cite{Takagi}
in the migrating direction.
Thus, we assume that the characteristic time $\tau_c$, during which the polarity direction persists,
is also much longer than the correlation time $\tau_R$ of the estimated direction.
We consider the integral of the driving force 
over time interval $\Delta t \gg \tau_R$,
$\boldsymbol{\Delta W}^\textrm{ext.}=\int_t^{t+\Delta t}\boldsymbol{\xi}^\mathrm{ext.}(s)ds$.
Then, by the central limit theorem, $\boldsymbol{\Delta W}^\textrm{ext.}$ follows the Gaussian distribution.
The average and mean square displacement 
of $\boldsymbol{\Delta W}^{\rm ext.}$ are given by 
\begin{equation} \label{eq:paverage}
 \langle\boldsymbol{\Delta W}^{\rm ext.}\rangle=\left( 0, \frac{f_q A}{2} \Delta t \right) \ .
\end{equation}
and 
\begin{equation} 
\label{eq:xiMSD}
 \langle |\boldsymbol{\Delta W}^\mathrm{ext.}|^2 \rangle
 = 2 f_q^2 \tau_R \Delta t.
\end{equation}
The driving force $S$ in Eq.~\eqref{eq:force} is given by the $y$-component of $\langle\boldsymbol{\Delta W}^{\rm ext.}\rangle$ in Eq. \eqref{eq:paverage} (divided by $\Delta t$).
Next, consider the change in the direction of internal polarity within $\Delta t$, 
$\Delta{\theta_q} = \theta_q(t+\Delta t) - \theta_q(t)$.
From Eq. \eqref{eq:tensorq} with $I_q \rightarrow \infty$, Eqs. \eqref{eq:xiintdispersion}, \eqref{eq:paverage}, and \eqref{eq:xiMSD},
the moments of $\Delta{\theta_q}$ up to $O(\Delta t)$ are given by
\begin{gather}
 \langle\Delta{\theta_q}\rangle = - \frac{1}{2} f_q A
 \sin {\theta_q}(t) \Delta t \label{eq:Moment1} \\
 \langle(\Delta{\theta_q})^2\rangle = (f_q^2 \tau_R + 2 D^{\rm int.}) \Delta t \label{eq:Moment2} \ .
\end{gather}
Using the Kramers-Moyal expansion with Eqs. \eqref{eq:Moment1} and \eqref{eq:Moment2},
we obtain the following Fokker-Planck equation
\begin{equation} \label{eq:Fokker-PlanckNeo}
 \frac{\partial}{\partial t} P(\theta_q,t) = 
   \frac{\partial}{\partial \theta_q} \left[ c_q(\theta_q) P(\theta_q,t) \right]
   + D \frac{\partial^2}{\partial \theta_q^2} P(\theta_q,t) \ ,
\end{equation}
where $c_q(\theta_q) = (f_q A \sin \theta_q)/2$,
and $D$ is the diffusion constant $D = D^{\rm ext.} + D^{\rm int.}$,
which consists of two independent dispersions $D^{\rm ext.}=f_q^2 \tau_R / 2$ and $D^{\rm int.}$,
respectively denoting the diffusion strengths introduced by external and internal perturbations.

\subsection*{Propagation of noise in the linear cascade reaction}

Here, we consider the contribution of the receptor noise 
to the downstream signals. 
Consider a linear signaling cascade, in which the activity of each reaction step is modulated by the one-upstream step. 
Then, the stochastic temporal evolutions of the small deviations $\eta_i$ from their stationary average at each step
are described by the following linearized Langevin equations~\cite{Shibata2005};
\begin{equation}
  \tau_i \frac{d}{dt} \eta_i = - \eta_i + \eta_{i-1} + \xi_i(t) \ ,
  \label{linearLangevin-i}
\end{equation}
where $\tau_i$ is the time constant (given by the inverse depletion rate) for each reaction step, 
and the dimensionless quantity $\eta_i$
includes signal amplification effects.
The noise in the activated receptor is given by $\eta_0$,
with $\langle \eta_0(t') \eta_0(t) \rangle = \sigma_0^2 \exp(-|t'-t|/\tau_R)$.
The last term $\xi_i$ denotes the noise in the $i$-th signal transduction.

From Eq. \eqref{linearLangevin-i}, the autocorrelation function is given by
\begin{equation} \label{eq:DXndispersion}
 \langle  \eta_n(t')  \eta_n(t) \rangle 
 = 
 \langle  \eta_n(t')  \eta_n(t) \rangle_0 
  +\Delta \ ,
\end{equation}
where the first term on the right hand side 
denotes the contribution of the noise in the receptor signal, $\eta_0$, to the noise at the $n$-th step of the signal cascade, given by 
\begin{equation} \label{eq:noisefromR}
 \langle  \eta_n(t')  \eta_n(t) \rangle_0 
  = \sigma_0^2 \Lambda _n (t'-t) \ .
\end{equation}
The function $\Lambda_n (\Delta t)$ describes the time-dependence of the receptor noise in $\eta_n(t)$:
\begin{equation}
\Lambda_n (\Delta t)
=
\sum_{i=0}^n \frac{\tau_R}{\tau_i} 
\left( \prod_{j=0,j\not=i}^n
\frac{\tau_i^2}{\tau_i^2-\tau_j^2} \right)
e^{-|\Delta t|/\tau_i}
\label{eq:intro-H}
\end{equation}
with $\tau_0\equiv\tau_R$.
The second noise term $\Delta$ on the right hand side of Eq. \eqref{eq:DXndispersion}
results from the signal cascade, and sums the various noises in the downstream signals, $\xi_i$.
Since $\Delta$ is included in
$D^{\rm int.}$ but not in $D^\textrm{ext.}$,
we focus on the receptor noise given by Eq. \eqref{eq:noisefromR}.

In {\it Dictyostelium} cells,
the receptor time constant $\tau_R$ is about $1$ second, while the
time constants $\tau_i$ of cascade reactions ($i=1,2,\cdots,n$) are comparable to or less than several seconds \cite{Cai2012,Kortholt2011}. On the other hand,
the persistence time of the migration direction is about $\tau_c \sim 300$ seconds.
Because
the time constants $\tau_i\quad(i=0,1,2\cdots,n)$ are much smaller than $\tau_c$,
we can replace $e^{-|\Delta t|/\tau_i}$ by $2 \tau_i \delta(\Delta t)$. Thus, we obtain
\footnote{The first equality in Eq. \eqref{eq:intro-Happ} is obtained by
the following relation;
$\sum_{i=0}^n (-1)^{i} A_i^n \prod_{j\neq i,k \neq i (j>k)} (A_j-A_k) = \prod_{j,k (j>k)}(A_j-A_k)$
for arbitrary numbers $A_i$ ($i=0,1,2\cdots,n$).
}
\begin{equation}
\Lambda_n (\Delta t)
= 2 \tau_R \delta (\Delta t) \sim e^{-|\Delta t|/\tau_R} \ ,
\label{eq:intro-Happ}
\end{equation}
and hence 
 $\langle  \eta_n(t')  \eta_n(t) \rangle_0 =
\langle  \eta_0(t')  \eta_0(t) \rangle$.
Therefore, the time constant $\tau_R (=\tau_0)$ can be used in Eqs. \eqref{eq:D_ext} and \eqref{eq:xiMSD}.
Intuitively, this argument implies that
the time constant $\tau_R$, considered as the interval of signaling events at the receptor, is not affected by the time delay between a particular signaling event 
and the resulting directional changes along the signaling cascades~(Fig.~\ref{fig:introduction}D).

\subsection*{Cell motile behavior in the uniform chemoattractant solution.}

The persistence of migration direction is characterized by 
the autocorrelation function of the migration direction. 
In the presence of a uniform chemoattractant 
with $C_0 > 0 $ and $A=0$,
the autocorrelation function of $\boldsymbol{q}(t)$,
$C(t)=\langle {\bm q}(t) \cdot {\bm q}(0) \rangle = \langle \cos( \theta_q(t)-\theta_q(0) ) \rangle$
is obtained from Eq.~\eqref{eq:Fokker-PlanckNeo} with $A=0$ as
\begin{align}
C(t) &= \int_0^{2\pi}   \cos \theta_q P(\theta_q,t|\theta_q=0,t=0)d\theta_q \notag \\
     &= \exp \left(-\frac{|t|}{D^{-1}} \right) \ , \label{eq:autocorrelation}
\end{align}
where the diffusion constant $D = D^{\rm ext.} + D^{\rm int.}$. 
In this manner, we obtain the correlation time $\tau_c$ in Eq.~\eqref{eq:tauvtauq}.

\subsection*{Stationary distribution of the migration direction and chemotaxis index in a shallow chemical gradient.}

Solving Eq. \eqref{eq:Fokker-PlanckNeo},
the stationary probability distribution $P_s(\theta_q)$ of the polarity and migration direction $\theta_v=\theta_q$ 
is given by the circular normal distribution as
\begin{equation} \label{eq:stationarydistribution} 
 P_s(\theta_q) = \frac{1}{I_0(\kappa)} \exp(\kappa \cos \theta_q) \ ,
\end{equation}
where the accuracy $\kappa$ is obtained from Eq. \eqref{eq:kappa}.
Here, $I_0(\kappa)$ is the modified Bessel function, defined as $I_0(\kappa) = \int d\theta_q \exp(\kappa \cos \theta_q)$. 
Eq. \eqref{eq:stationarydistribution} is plotted as a function of migration distance in Fig.~\ref{fig:chemotaxisindex}A (red solid line). The parameter values of a {\it Dictyostelium} cell are assumed.
From Eq. \eqref{eq:stationarydistribution},
the chemotaxis index is obtained as 
\begin{equation} \label{eq:CIresult}
 {\rm CI} \equiv \int d\theta_q \cos \theta_q P_s(\theta_q) 
= \frac{\partial}{\partial \kappa} \log I_0(\kappa)\Bigg|_{\kappa=(f_q A)/(2 D_q)} \ .
\end{equation}
The chemotaxis index ${\rm CI}$, given by Eq. \eqref{eq:CIresult}, saturates at $1$; that is,
${\rm CI} \sim 1 - (f_q \tau_R + 2 f_q^{-1} D^{\rm int.})/(2 A) \rightarrow 1 $
as $A \rightarrow \infty$, whereas 
${\rm CI} \sim A/(2 f_q \tau_R + 4 f_q^{-1} D^{\rm int.}) = \kappa/2$
as $A \rightarrow 0$.
Figure~\ref{fig:chemotaxisindex}C plots
the chemotaxis index \eqref{eq:CIresult} with Eqs. \eqref{eq:A0extapprox} and \eqref{eq:muwithout} (red solid line) as functions of chemoattractant concentration assuming the
experimental parameters $k_d=1 \ {\rm s}^{-1}$ and $K_d=180 \ {\rm nM}$ and 
the parameter values \eqref{eq:fittingfq} and \eqref{eq:fittingDI}.

\subsection*{Gradient sensing in cell}

Here, we extend the maximum likelihood estimates of gradient direction reported in Refs. \cite{Hu2010PRL,BabaPre}
to include the effect of stochasticity at the level of the second messenger~\cite{Ueda08,Amselem2012}.
We denote the concentrations of activated receptor and 
the activated second messenger at $\theta$
by $r^*(\theta)$ and $x^*(\theta)$, respectively.
Under an exponential chemoattractant gradient of steepness~$p$,
the chemoattractant concentration along the periphery of the cell is given by $C(\theta)=C_0\exp{((p/2)\cos(\theta-\phi))}$. 
When cells estimate the gradient direction from the distribution of the activated second messenger, the most probable direction $\phi$ is given by
$s\cos{\phi}=Z_1$ and $s\sin{\phi}=Z_2$, where $s^2=Z_1^2+Z_2^2$ and 
\begin{eqnarray}
Z_1&=&\int_0^{2\pi} x^{*}(\theta) \cos \theta d\theta \ , \\
Z_2&=&\int_0^{2\pi} x^{*}(\theta) \sin \theta d\theta \ .
\end{eqnarray}
The distribution of the estimated $\phi$
can be derived from the distributions of $Z_1$ and $Z_2$.
Here, we consider the distribution of $\phi$ up to first order in $p$, from which we calculate the average and the dispersion of $x^{*}(\theta)$.
The averages of $r^{*}$ and $x^{*}$ are obtained as
\begin{align}
\langle r^*(\theta)\rangle &=
\frac{C(\theta)}{C(\theta)+K_d}r_0 \ , \\
\langle x^*(\theta)\rangle &=
\frac{\langle r^*(\theta)\rangle}{\langle r^*(\theta)\rangle+K_d}x_0 \notag \\
&= \frac{C(\theta)r_0}{C(\theta)r_0+K_\mathrm{x}(C(\theta)+K_d)}x_0 \ ,
\end{align}
where $r_0=N/2\pi$, $x_0=X_t/2\pi$ and $K_\mathrm{x}=k_\mathrm{xd}/k_\mathrm{xp}$,
and $N$ and $X_t$ are the total concentrations of receptor and second messenger, respectively. The production and degradation rates of the second messenger are ~$k_\mathrm{xp}$ and $k_\mathrm{xd}$, respectively.
The fluctuation densities of $r^*(\theta)$ and $x^*(\theta)$,
defined by
$\langle r^*(\theta)r^*(\theta')\rangle=\sigma_r^2(\theta)\delta(\theta-\theta')$
and
$\langle x^*(\theta)x^*(\theta')\rangle=\sigma_x^2(\theta)\delta(\theta-\theta')$,
are given by
\begin{align}
\sigma_r^2(\theta)&=
\frac{C(\theta)K_d}{\left(C(\theta)+K_d\right)^2}r_0 \ , \\
\sigma_x^2(\theta)
&=g_\mathrm{X}(\theta)\langle x^*(\theta)\rangle \notag \\
&+g_\mathrm{X}(\theta)^2\frac{\tau_R(\theta)}{\tau_R(\theta)+\tau_X(\theta)}\frac{\sigma_r^2}{\langle r^{*}\rangle^2}\langle x^*(\theta)\rangle^2 \ ,
\end{align}
where
$g_\mathrm{x}(\theta)=K_X/(K_X+r^*)$,
$\tau_R(\theta)=(k_{\mathrm{on}}C(\theta)+k_{\mathrm{off}})^{-1}$
and
$\tau_x(\theta)=(k_\mathrm{xp}r^*(\theta)+k_\mathrm{xd})^{-1}$.
Using the average and dispersion of $x^*(\theta)$,
the distribution of $\phi$ was evaluated up to first order of $p$ (see Eq. \eqref{eq:Pphi1})
with 
\begin{equation} \label{eq:A0extapprox}
 A = \frac{1}{2} p \pi^{1/2} \mu^{1/2},
\end{equation}
and 
\begin{align}
 \mu^{-1} 
 &=\frac{2 (C_0/K_d+1)^2 }{\pi C_0 K_X r_0 x_0}
\Bigg\{ \frac{[K_d K_X + C_0(K_X+r_0)]^2}{C_0+K_d} \notag \\
&+\frac{K_d K_X [K_X(C_0+K_d) +C_0 r_0]x_0}{(C_0+K_d)[K_X+(K_d+C_0)k_{\textrm{on}}/k_{\textrm{xp}}]+C_0 r_0}
\Bigg\}. \label{eq:muwith}
\end{align}
Taking $x_0 \rightarrow \infty$, followed by the limit $k_{\textrm{xp}} \rightarrow \infty$,
we have $\mu = \mu_{\rm mee.}$ where
\begin{equation} \label{eq:muwithout}
 \mu_{\rm mee.} \equiv \frac{C_0 K_d N}{4 (C_0+K_d)^2} \ .
\end{equation}
This form of $\mu_{\rm mee.}$ is consistent with the result 
based on the most efficient estimation (maximum likelihood estimate without
stochasticity in the second messenger reaction) \cite{BabaPre}.
We also note that, as $C_0 \rightarrow \infty$, we have 
$\mu \sim 2 \pi K_X K_d^2 r_0 x_0 C_0^{-2} /(K_X+r_0)^2 \propto C_0^{-2}$.
Thus, this function converges much faster than $\mu_{\rm mee.}$ derived from the most efficient estimate, 
which decreases as $\propto C_0^{-1}$ as $C_0 \rightarrow \infty$.
Moreover, the coefficient is proportional to the second messenger concentration $x_0$.
Therefore, consistent with a previous report \cite{Ueda08},
the activation of the second messenger becomes the bottleneck of the chemotactic 
signal processing at high chemoattractant concentrations.

\subsection*{Cell motility assay}

\textit{Dictyostelium discoideum} AX2 cells (wild type) 
were starved by suspension in development buffer (DB : 5 mM Na phosphate buffer, 2 mM MgSO$_4$, 0.2 mM CaCl$_2$, pH 6.3) 
for 1 hour and were then pulsed with 10 nM cAMP at 6-minute intervals for up to 3.5 hours at 21 $^\circ$C, 
resulting in elongated cells  with chemotactic competency \cite{Sato07}.
The prepared cells were settled on a glass dish (IWAKI) at cell density 
$1.0\times10^5$ cell/ml (to preclude explicit cell-cell interactions), containing DB supplemented with a given concentration of cAMP and
2 mM dithiothreitol (DTT, Sigma) was added to inhibit phosphodiesterase in order to keep the cAMP concentration in the medium. Cells were incubated under these conditions for 20 min. 
Phase-contrast imaging was performed using an inverted microscopes (TiE, Nikon) 
with a 40$\times$ phase-contrast objective, equipped with an EMCCD camera (iXon+, Andor). 
Images of cells were taken every second for 30 min.
The cell periphery of individual cells was detected
and the trajectories of cell area centroids were obtained
by a customized program developed in Matlab 7.6 (Mathworks). 
All trajectories were analyzed for longer than 5 min. 
The cell velocity $\boldsymbol{v}_n$ in frame $n$ was obtained as
$\boldsymbol{v}_n=\boldsymbol{r}_{n+1}-\boldsymbol{r}_n$, where
~$\boldsymbol{r}_n$ denotes the centroid trajectory.

\section*{Acknowledgments}
This work has been supported by KAKENHI (25111736).
We are grateful to Dr. A. Baba for his contribution at the early stage of this work and Dr. K. Sato for helpful comments.


\end{document}